\documentclass[12pt]{article}

\usepackage{epsfig}
\usepackage{amssymb}

\makeatletter
   \newcommand\tabcaption{\def\@captype{table}\caption}
\makeatother

\newcommand{\be}{\begin{equation}}
\newcommand{\ee}{\end{equation}}
\newcommand{\bea}{\begin{eqnarray}}
\newcommand{\eea}{\end{eqnarray}}

\newcommand{\Rh}{{\mathbb R}}
\newcommand{\Dc}{{\mathcal D}}

\newcommand{\Nc}{{\mathcal N}}

\newcommand{\pd}{\partial}

\begin{document}
\title{
\begin{flushright}
{\small SMI-5-00 }
\end{flushright}
\vspace{1cm}UV/IR Mixing for\\
Noncommutative Complex Scalar Field Theory, II\\
(Interaction with Gauge Fields)}
\author{
I. Ya. Aref'eva${}^{\S}$, D. M. Belov${}^{\dag}$,
A. S. Koshelev${}^{\dag}$ and O. A. Rytchkov${}^{\S}$\\
\\${}^{\S}$
{\it  Steklov Mathematical Institute,}\\ {\it Gubkin str.8,
Moscow, Russia, 117966}\\ arefeva@genesis.mi.ras.ru, rytchkov@mi.ras.ru\\
\\${}^{\dag}$
{\it Physical Department, Moscow State University, }\\ {\it
Moscow, Russia, 119899} \\ belov@orc.ru, kas@depni.npi.msu.su}

\date {~}
\maketitle
\begin{abstract}
We consider noncommutative analogs of scalar electrodynamics and
$\Nc =2$ $D=4$ SUSY Yang-Mills theory.
We show that one-loop renormalizability of
noncommutative scalar electrodynamics
requires the scalar potential to be an anticommutator squared.
This form of the scalar potential differs from the one expected
from the point of view of noncommutative
gauge theories with extended SUSY
containing a square of commutator. We show that fermion contributions
restore the commutator in the scalar potential.
This provides one-loop
renormalizability of noncommutative $\Nc =2$ SUSY gauge theory.
We demonstrate a presence of non-integrable IR singularities in
noncommutative scalar electrodynamics for general coupling constants.
We find that for a special ratio of coupling constants these IR
singularities vanish. Also we show that IR poles are absent
in noncommutative $\Nc =2$ SUSY gauge theory.
\end{abstract}

\newpage
\section{Introduction}

Recently, there is a renovation of the interest in noncommutative
quantum field theories (or field theories on noncommutative
space-time \cite{book,Mad}). As emphasized in \cite{SW}, the
important question is whether or not the noncommutative quantum
field theory  is well-defined. Note that one of earlier
motivations to consider noncommutative field  theories  is a hope
that it would be possible to avoid  quantum field theory
divergencies \cite{WZ,AVqp,Mad,filk,Kempf,AV}. One could think
that a theory on a noncommutative space is
renormalizable only if the corresponding commutative theory is
renormalizable. Results on one-loop renormalizability of
noncommutative gauge theory \cite{ren} and   two-loop
renormalizability of noncommutative scalar $\phi _4^4$ theory
\cite{ABK} as well as general considerations \cite{SB,Ch} support
this belief.
But this expectation is not true for more complicated
models. In \cite{ABK2} a noncommutative
quantum field theory of a complex scalar field
with the interaction
$\lambda^2(a\phi^*\star\phi\star\phi^*\star\phi+
b\phi^*\star\phi^*\star\phi\star\phi)$
has been considered. Here $\lambda$ is a coupling constant.
It has been shown that this model is renormalisable at one-loop in two
special cases: $a=b$ and $b=0$.

In this paper we analyze an interaction of the noncommutative complex
scalar field with the noncommutative $U(1)$ gauge field
(noncommutative scalar electrodynamics). We argue that one-loop
renormalizability requires the scalar
potential to be an anticommutator squared.
However, $\Nc=2$
supersymmetric Yang-Mills theory \cite{booksuper,zum} contains
a complex scalar field with the potential which is equal to the
square of the commutator. At
the first sight it seems that the above fact contradicts to
expected renormalisability of noncommutative $\Nc=2$
SUSY Yang-Mills theory \cite{Sus,FL}. Nevertheless, it  turns out that
fermions remove this discrepancy. Namely, fermion contributions
restore a commutator in the scalar potential.
Thus our analysis confirms one-loop renormalisability of  noncommutative
$\Nc=2$ SUSY Yang-Mills theory.

Note that UV renormalizability does not guarantee that the
theory is well-defined. There is a mixing of the UV and IR
divergencies \cite{MRS,ABK}. UV/IR mixing depends on the model \cite{ABK2,Sus}.
The $U(1)$ noncommutative gauge theory does not
exhibit a mixing
of the UV and the IR dynamics \cite{Hay}.
For further developments in perturbative study of
noncommutative field theories see
\cite{AV-M}-\cite{CK}, for non-perturbative aspects see \cite{IKK}-\cite{GMS}.
In this paper we show that one can remove
IR divergencies in
noncommutative scalar electrodynamics in the case of a special relation
between two coupling constants.
Also we show that IR poles are absent
in noncommutative $\Nc =2$ SUSY gauge theory.

The paper is organized as follows. In Section $2$ we analyze
noncommutative scalar electrodynamics. In Section $3$ we consider
noncommutative $\Nc=2$ SUSY Yang-Mills theory (NC SYM)
and prove its one-loop renormalizability. In Section 4 we analyze
infrared behaviour of noncommutative  scalar electrodynamics and $\Nc=2$ NC SYM.
We show that the latter theory is well defined meanwhile the first
one requires a
special relation between two coupling constants.

\section{Scalar Electrodynamics}
Let us start with a consideration of noncommutative scalar
electrodynamics in Euclidean space $\Rh^4$. The classical action
is given by
\bea &S=\int d^4x\left(-\frac{1}{4} F_{\mu\nu}\star F^{\mu\nu}
+(\Dc_{\mu}\phi^*)\star
(\Dc_{\mu}\phi)+V[\phi^*,\phi]\right),&\nonumber
\label{action1}\\
&V[\phi^*,\phi]=\lambda^2(a\phi^*\star\phi\star\phi^*\star\phi+
b\phi^*\star\phi^*\star\phi\star\phi),&
\eea
where $\star$ is a
Moyal product $(f\star g)(x)=\exp(-i\xi
\theta^{\mu\nu}\pd_{\mu}\otimes\pd_{\nu})f(x)\otimes g(x)$ with
$\theta^{\mu\nu}$ being a real constant skew-symmetric
nondegenerate matrix and $\xi$ being a deformation parameter. The
covariant derivative is defined by $\Dc_{\mu}\phi=\pd_{\mu}\phi-
ig[A_{\mu},\phi]_{\star}$,
where $[A_{\mu},\phi]_{\star}=A_{\mu}\star\phi-\phi\star A_{\mu}$.
$g$ and $\lambda$ are coupling
constants and $a$ and $b$ are fixed real numbers. It has been shown in
\cite{ABK2}, that the pure complex scalar field theory is one-loop
renormalizable only if $a=b$ or $b=0$. The purpose of the present
analysis is to find the analogous restrictions on $a$ and $b$ in
the case of scalar electrodynamics.

The action (\ref{action1}) is invariant under the following gauge
transformations
\be
\phi\mapsto U\star\phi\star U^{\dag},\quad
\phi^*\mapsto U\star\phi^*\star U^{\dag},\quad A_{\mu}\mapsto
U\star A_{\mu}\star U^{\dag}-\pd_{\mu}U\star U^{\dag},
\ee where
$U$ is an element of the noncommutative $U(1)$ group \cite{FL}.
Note that since our fields are in adjoint representation we could consider
the theory with two real scalar fields instead of one complex.

The Feynman rules for the theory (\ref{action1}) are presented in
Table 1. Solid lines denote the scalar fields, "in" arrows
stand for the field $\phi$ and "out" arrows stand for the field
$\phi^*$.

\begin{center}
\renewcommand{\arraystretch}{2.5}
\begin{tabular}{cl}
\epsfig{file=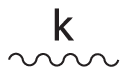,
   width=30pt,
  angle=0,
 }
 & $~~~~~D_{\mu\nu}(k)=-\frac{1}{k^2}\left[\delta_{\mu\nu}-
 (1-\alpha)\frac{k_{\mu}k_{\nu}}{k^2}\right]$ \\
\epsfig{file=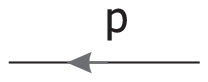,
   width=40pt,
  angle=0,
 }
 & $~~~~~D(p)=\frac{1}{p^2}$ \\
\epsfig{file=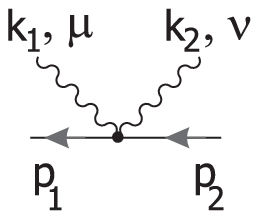,
   width=50pt,
  angle=0,
 }
 & $~~~~~4g^2\delta^{\mu\nu}
 [\cos(k_1\wedge p_1+k_2\wedge p_2)-\cos(p_1\wedge p_2)\cos(k_1\wedge k_2)]$ \\
\epsfig{file=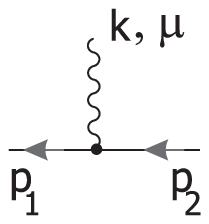,
   width=40pt,
  angle=0,
 }
 & $~~~~~2ig(p_1-p_2)_{\mu}\sin(p_1\wedge p_2)$ \\
\epsfig{file=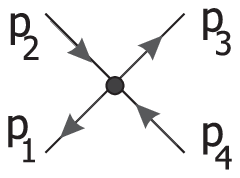,
   width=40pt,
  angle=0,
 }
 & $~~~~~-4\lambda^2[a\cos(p_1\wedge p_2+p_3\wedge p_4)+b\cos(p_1\wedge p_3)
 \cos(p_2\wedge p_4)]$ \\
\epsfig{file=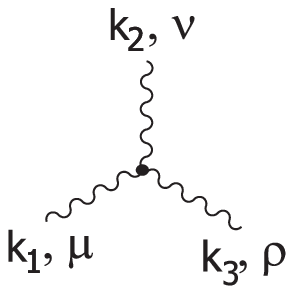,
   width=50pt,
  angle=0,
 }
 & $~~~~~-2ig\sin(k_1\wedge k_2)[
 (k_1-k_2)_{\rho}\delta_{\mu\nu}+
 (k_2-k_3)_{\mu}\delta_{\nu\rho}+(k_3-k_1)_{\nu}\delta_{\mu\rho}]$ \\
\end{tabular}

\begin{minipage}[b]{54pt}
\centering \epsfig{file=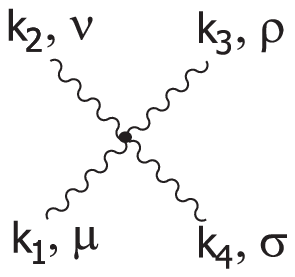, width=50pt, angle=0,}
\end{minipage}
\begin{minipage}[b]{.75\textwidth}
\raggedright
\bea
4g^2[(\delta_{\mu\rho}\delta_{\nu\sigma}-\delta_{\mu\sigma}\delta_{\nu\rho})
\sin (k_1\wedge k_2)\sin (k_3\wedge k_4)\nonumber\\
+(\delta_{\mu\nu}\delta_{\rho\sigma}-\delta_{\mu\sigma}\delta_{\nu\rho})
\sin (k_1\wedge k_3)\sin (k_2\wedge k_4)\nonumber\\
+(\delta_{\mu\nu}\delta_{\rho\sigma}-\delta_{\mu\rho}\delta_{\nu\sigma})
\sin (k_1\wedge k_4)\sin (k_2\wedge k_3)]\nonumber
\eea
\end{minipage}
\tabcaption{Feynman rules for scalar electrodynamics.}
\end{center}

We do not specify the Feynman rules for ghosts, since they do not
contribute to one-loop graphs with the external matter lines.

All calculations are performed in Landau gauge $\alpha=0$. This is
a convenient choice, since in this gauge a great number of
graphs do not have divergent parts. One can prove that the theory
is gauge invariant on
quantum level (cf. \cite{ren,Sus}), so our results and conclusions are
valid for an arbitrary value of $\alpha$.

\begin{figure}[h]
\begin{center}
\epsfig{file=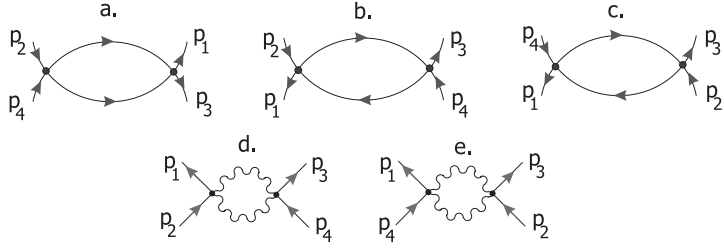, width=270pt, angle=0, } \caption{One-loop
corrections to the $4$-point scalar vertex.} \label{F1}
\end{center}
\end{figure}

The above mentioned restriction on $a$ and $b$
can come from one-loop corrections to the 4-scalar and
2-scalar-2-gluon vertices.
First we consider one-loop corrections to the $4$-point scalar
vertex. The graphs that have non zero divergent parts in Landau
gauge are presented in Figure \ref{F1}.
Using the dimensional regularization
($D=4-2\epsilon$)
we find that the sum of divergent parts of these graphs is equal to
\be
\frac{4}{(4\pi)^2\epsilon}
[(3g^4+4\lambda^4a^2+\lambda^4b^2)
\cos(p_1\wedge p_2+p_3\wedge p_4)+
(3g^4+4\lambda^4ab+\lambda^4b^2)
\cos(p_1\wedge p_3)\cos(p_2\wedge p_4)].
\label{4sb}
\ee
The condition of one-loop renormalizability yields a system of
two algebraic equations on $a$ and $b$
\bea
3g^4+4\lambda^4a^2+\lambda^4b^2&=&ca,\\
3g^4+4\lambda^4ab+\lambda^4b^2&=&cb,
\eea
where $c$ is a constant. These equations are self-consistent only in
the case $a=b$. Therefore the renormalizable potential for scalar
electrodynamics has the form
\be
V[\phi^*,\phi]=a\frac{\lambda^2}{2}(\{\phi^*,\,\phi\}_{\star})^2,
\label{potential}
\ee
where $\{f,\,g\}_{\star}=f\star g+g\star f$.
Note that in contrast to the pure noncommutative complex scalar field
theory \cite{ABK2} we do not have the solution $b=0$.

\begin{figure}[h]
\begin{center}
\epsfig{file=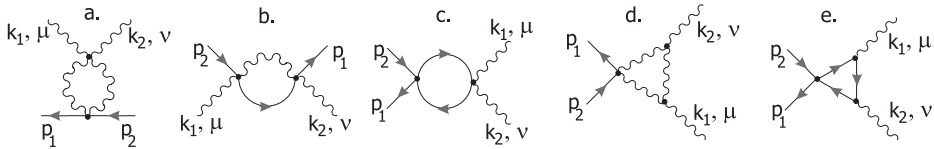,width=350pt, angle=0,} \caption{One-loop
corrections to the $2$-scalar-$2$-gluon vertex.} \label{F2}
\end{center}
\end{figure}

Let us turn to an analysis of one-loop corrections to the
$2$-scalar-$2$-gluon vertex. The graphs that have non zero
divergent parts in the $\alpha=0$ gauge are presented in Figure
\ref{F2}.

The sum of the divergent parts of these graphs is
\bea
\frac{12}{(4\pi)^2\epsilon}
g^4\delta_{\mu\nu}[\cos(p_1\wedge k_1+p_2\wedge
k_2)-\cos(p_1\wedge p_2)\cos(k_1\wedge k_2)].
\label{2s2gb}
\eea
Note that the graphs \ref{F2}c and \ref{F2}e do contain $4$-point
scalar vertex and there are terms depending on $a$ and $b$.
However the contributions of these graphs mutually cancel and the
sum does not depend on $a$ and $b$. Therefore, there are no
new restrictions on these constants. We see that a counterterm
requiring for a cancellation of (\ref{2s2gb}) has just the same
trigonometric structure as the initial $2$-scalar-$2$-gluon
vertex in the action, i.e. this vertex is one-loop renormalizable.

Thus the above analysis leads to the conclusion that noncommutative
scalar
electrodynamics (\ref{action1}) is one-loop renormalizable only if
the scalar potential has the anticommutator form (\ref{potential}).

\section{Noncommutative $\Nc=2$ Super Yang-Mills Theory.}

The action for the
Euclidean noncommutative $\Nc=2$ SUSY Yang-Mills theory reads
\bea
&S=\int d^4x\left(-\frac{1}{4} F_{\mu\nu}\star F^{\mu\nu}
+(\Dc_{\mu}\phi_-)\star (\Dc_{\mu}\phi_+)
-i\chi^*\star\widehat{\Dc}\chi\right.&\nonumber\\
&\left.-g\sqrt{2}\chi^*\star(R[\chi,\phi_+]_{\star}+L[\chi,\phi_-]_{\star})
-\frac{g^2}{2}([\phi_-,\phi_+]_{\star})^2\right),& \label{action2}
\eea where
$\Dc_{\mu}=\pd_{\mu}-ig[A_{\mu},\cdot]_{\star}$,
$L,R=\frac12(1\pm\Gamma^5)$, $\phi_{\pm}$ are real scalar fields,
$\chi$ is a complex four-component spinor\footnote{Note that $\Nc=2$
SYM in Minkowski space contains a complex scalar field.}. The action
(\ref{action2}) is a noncommutative generalization of Euclidean
$\Nc=2$ SYM theory \cite{zum}. A formulation of noncommutative
$\Nc=2$ supersymmetric theories in terms of superfields was
given in \cite{FL}.

Note that the scalar
electrodynamics examined in the previous section can be
considered as a bosonic part of $\Nc=2$ NCSYM. The identification
is evident: $\phi$ and $\phi^*$ corresponds to $\phi_+$ and
$\phi_-$, respectively.
Also we should replace $\lambda$ by $g$ and take $a=-b=-1$. The
fact that fields $\phi_{\pm}$ are not complex
conjugate does not affect the performed calculations. The
Feynman rules for the bosonic part of the action
(\ref{action2}) can be easily obtained from the Feynman rules for
scalar electrodynamics (see Table 1) using
the above mentioned identification. The Feynman rules for fermion
fields are presented in Table 2.

\renewcommand{\arraystretch}{2.5}
\begin{minipage}{0.5\textwidth}
\begin{tabular}{cl}
\epsfig{file=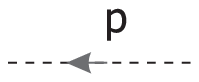,
   width=40pt,
  angle=0,
 }
 & $~~~~~S(p)=\frac{\Gamma^{\mu}p_{\mu}}{p^2}$ \\
\epsfig{file=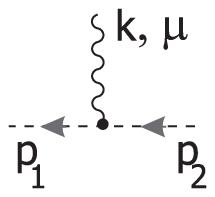,
   width=40pt,
  angle=0,
 }
 & $~~~~~-2ig\Gamma^{\mu}\sin(p_1\wedge p_2)$ \\
\end{tabular}
\end{minipage}
\begin{minipage}{0.5\textwidth}
\begin{tabular}{cl}
\epsfig{file=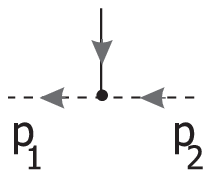,
   width=40pt,
  angle=0,
 }
 & $~~~~~2ig\sqrt{2}\sin(p_1\wedge p_2) R$ \\
\epsfig{file=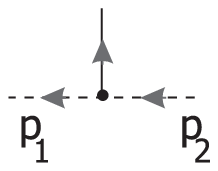,
   width=40pt,
  angle=0,
 }
 & $~~~~~2ig\sqrt{2}\sin(p_1\wedge p_2) L$ \\
\end{tabular}
\end{minipage}
\tabcaption{Feynman rules for fermion fields.}

\begin{figure}[t]
\begin{center}
\epsfig{file=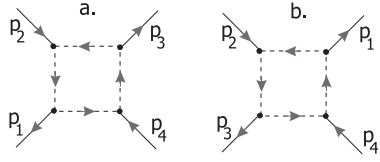, width=140pt, angle=0, }
\caption{One-loop fermion corrections to the $4$-point scalar vertex.}
\label{F3}
\end{center}
\end{figure}

As in the case of scalar electrodynamics we start with the
examination of the one-loop corrections to the $4$-point scalar
vertex. The graphs with fermion loops are presented in Figure
\ref{F3}.
The divergencies coming from these graphs are
\be
-\frac{32}{(4\pi)^2\epsilon}
g^4\cos(p_1\wedge p_2+p_3\wedge p_4).
\label{4sf}
\ee
Thus, taking into account the contribution (\ref{4sb}) of the
boson graphs we find that the boson and fermion divergencies
mutually cancel.
The similar result is valid for ordinary $\Nc=2$ SUSY Yang-Mills
theory where
the 4-point scalar vertex is finite at one-loop \cite{booksuper}.

\begin{figure}[t]
\begin{center}
\epsfig{file=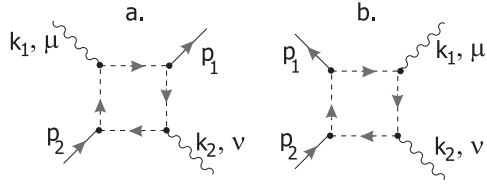,width=180pt, angle=0,} \caption{One-loop
fermion corrections to the $2$-scalar-$2$-gluon vertex.}
\label{F4}
\end{center}
\end{figure}

Next we calculate one-loop corrections to the $2$-scalar-$2$-gluon
vertex. The graphs with fermion loops are presented in Figure \ref{F4}.
The sum of divergent parts of these graphs is
\be
-\frac{16}{(4\pi)^2\epsilon}
g^4\delta_{\mu\nu}[\cos(p_1\wedge k_1+p_2\wedge
k_2)-\cos(p_1\wedge p_2)\cos(k_1\wedge k_2)].
\label{2s2gf}
\ee
Summing the contributions of the boson (\ref{2s2gb})
and fermion (\ref{2s2gf})
graphs we get
\be
-\frac{4}{(4\pi)^2\epsilon}
g^4\delta_{\mu\nu}[\cos(p_1\wedge k_1+p_2\wedge
k_2)-\cos(p_1\wedge p_2)\cos(k_1\wedge k_2)].
\ee
Note that one-loop fermion corrections as well as boson ones restore
the trigonometric structure of the initial vertex.
So, we conclude that the noncommutative $\Nc=2$ $D=4$ SYM with the
action (\ref{action2}) is one loop renormalizable.

\section{Infrared Behaviour.}

The finite parts of Feynman graphs contain the terms, which have
nonanalytic behaviour in $\xi$ and in external momenta. There is
a nontrivial mixing of UV and IR divergencies (see
\cite{MRS,ABK} for details).
It is straightforward to show that a type of the UV
divergency of the integral $J_{UV}(\xi p)=\int f(k,p)dk$
coincides with a type of the IR behaviour of
the integral $J_{IR}(\xi p)=\int e^{i\xi k\theta p}f(k,p)dk$.
For example, the logarithmically divergent integral $J_{UV}$ yields
a logarithmic singularity $\log(\xi|\theta p|)$ in $J_{IR}$,
the quadratically divergent integral $J_{UV}$ yields
a quadratic singularity $(\xi|\theta p|)^{-2}$ in $J_{IR}$,
and so on.
One has to care only about the poles, since near the origin the
logarithm is an integrable function. IR poles appear in the
corrections to propagators and can produce IR divergencies in
multi-loop graphs.

The one-loop correction to the gauge
propagator has been computed in \cite{Sus} and it has only
logarithmic IR singularity.
We examine one-loop corrections
to the scalar field propagator in the case of noncommutative
scalar electrodynamics (\ref{action1}) and the noncommutative $\Nc=2$
NC SYM theory (\ref{action2}). We do not analyze one-loop
corrections to the fermion propagator since they have only
logarithmic UV divergencies.

\begin{figure}[h]
\begin{center}
\epsfig{file=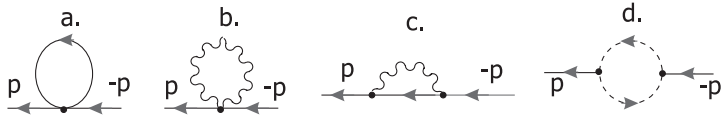,
   width=270pt,
  angle=0,
 }
\caption{One-loop corrections to the scalar field propagator}
\label{Fps}
\end{center}
\end{figure}

In the case of scalar electrodynamics all corrections to the
scalar field
propagator are presented in Figure \ref{Fps}a,b,c.
For $a=b$
the sum of the divergent parts is:
\be
\frac{6g^2}{(2\pi)^4}\int\frac{d^4k}{k^2}\left[1-\frac{a\lambda^2}{g^2}-
\left(1+\frac{a\lambda^2}{3g^2}\right)\cos(2k\wedge p)\right]-
\frac{8g^2}{(2\pi)^4}\int d^4k\frac{1-\cos(2k\wedge p)}{k^2(p+k)^2}
\left[p^2-\frac{(pk)^2}{k^2}\right]
\label{phipb}
\ee
where $p$ is an external momentum. A quadratic UV divergence
in (\ref{phipb}) is removed by a mass renormalization.
To remove IR poles one has to impose the condition $\lambda^2a=-3g^2$.

Next we compute one-loop correction to the scalar field
propagator in the $\Nc=2$ $D=4$ NC SYM (\ref{action2}). We
have one more graph presented in Figure \ref{Fps}d. Summing all
contributions we have
\be
-4g^2\int\frac{d^4k}{(2\pi)^4}\frac{1-\cos(2k\wedge p)}{k^2(p+k)^2}
\left[p^2+2\frac{(pk)^2}{k^2}\right]
\ee
It is remarkable, that all quadratically divergent integrals vanish.
Therefore, only the term with an integrable logarithmic IR
singularity appears after integration.

\section*{Acknowledgments}

We would like to thank P. B. Medvedev for useful discussions.
 This work was supported in part
by RFBR grant 99-01-00166,  by grant for the leading scientific
schools and INTAS grants 99-00545 and 99-00590.
O.R. was supported in part by INTAS grant 96-0457 within the
research program of the International Center for Fundamental Physics
in Moscow.

\end{document}